\documentclass{phb-proc4-auth}

\usepackage{graphicx}
\usepackage{amssymb}
\begin{document}
\begin{frontmatter}

\journal{SCES '04}

\title{Magnetism  in purple bronze Li$_{0.9}$Mo$_6$O$_{17}$}

\author[MPI]{J. Chakhalian\corauthref{1}} 
\author[TR]{Z. Salman}
\author[UBC]{J. Brewer}
\author[UBC]{A. Froese}
\author[OAK]{J. He}
\author[OAK]{D. Mandrus}
\author[OAK]{and R. Jin}

\address[MPI]{Max Planck Institute for Solid State Research, Heisenbergstr. 1, Stuttgart D-70569, Germany}
\address[TR]{TRIUMF 4004 Wesbrook Mall, Vancouver, B.C., Canada V6T 2A3}
\address[UBC]{he University of British Columbia,6224  Agricultural Road, Vancouver, V6T 1A1, Canada}
\address[OAK]{Oak Ridge National Laboratory, P.O. Box 2008, Oak Ridge, Tennessee 37831, USA}


\corauth[1]{Corresponding Author: tel.: +49 711 689-1731; fax: +49 711 689-1632; e-mail: j.chakhalian@fkf.mpg.de.}


\begin{abstract}
Muon spin relaxation measurements around the 25 K metal-insulator
transition in Li$_{0.9}$Mo$_6$O$_{17}$ elucidate a profound role of disorder  as a possible mechanism for this transition. The relaxation rate $1/T_1$ and the muon Knight shift  are incompatible with  the  transition to a SDW state and thus exclude it. 
\end{abstract}



\end{frontmatter}
Recent optical studies of the ternary molybdenum oxide bronzes
Li$_{0.9}$Mo$_6$O$_{17}$  have kindled new interest in the electronic properties of
inorganic quasi-$1d$ transition metal compounds.\cite{1} Li$_{0.9}$Mo$_6$O$_{17}$ (a
"purple bronze") exhibits a particularly interesting set of
electronic and magnetic properties. For instance, in a recent series
of ARPES measurements it has been suggested that the ground state for
this system is the Tomonaga-Luttinger liquid.\cite{4} Li$_{0.9}$Mo$_6$O$_{17}$ has a
monoclinic crystal structure [$P1 21/m 1(11)$] and a unit cell
consisting of four MoO$_6$ octahedra and two MoO$_4$ tetrahedra (see Fig.
1).  The crystal structure is formed by stacking layers ($a-b$ planes)
of corner sharing octahedra and tetrahedra. Li$_{0.9}$Mo$_6$O$_{17}$ has a highly
anisotropic metallic conductivity down to 25 K, where its resistivity
exhibits an upturn and resembles that of a semiconductor. At 2 K the
material undergoes a transition to a BCS superconducting state.
Despite extensive (and ongoing) studies of the transition at 25 K, its
nature is still rather puzzling. The following models have been
proposed as likely candidates: CDW, SDW and localization driven by
disorder.\cite{5}

Muon spin relaxation ($\mu$SR) is sensitive to magnetic transitions in a
large variety of systems, so we performed high transverse field (hTF)
and zero field (ZF) $\mu$SR experiments on Li$_{0.9}$Mo$_6$O$_{17}$ with particular
attention to the 25 K transition, to clarify its magnetic nature.
 \begin{figure}
     \centering
     \includegraphics[width=\columnwidth]{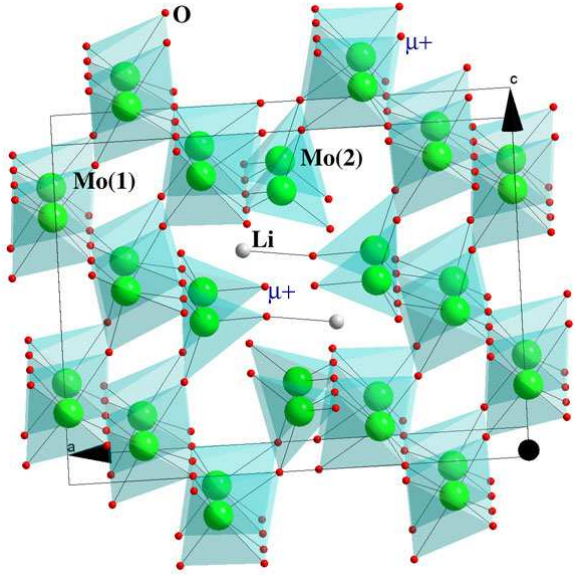}
     \caption{A purple bronze unit cell.  The molybdenum ions are located in an octahedral and 
a tetrahedral environment, respectively.  
Also shown are two possible $\mu^+$ sites bound to tetra- and octahedral oxygen ions.} 
 \end{figure}  

All the measurements were performed on the M15 beamline at TRIUMF,
which delivers nearly 100$\%$ spin polarized positive muons with a mean
momentum of 28 MeV/c. The muon spin polarization was rotated
perpendicular to the axis of the superconducting solenoid and muon
beam direction.  The magnitude of the applied magnetic field, $H=55$
kOe, was chosen to provide a balance between the magnitude of the
frequency shift, which increases with field, and the amplitude of the
$\mu$SR signal, which eventually diminishes with increasing field due to
the finite time resolution of the detectors.  The transverse field
precession measurements were all performed with a special cryostat
insert which allows high-field $\mu$SR spectra to be acquired in the
sample and in a reference material (CaCO$_3$) simultaneously.

Muon Knight shift data were taken on a powder sample of Li$_{0.9}$Mo$_6$O$_{17}$.
The magnitude of the shift was calculated as $K=(\omega_{\mu}-\omega_L)/\omega_L$,
where $\omega_{\mu}$ is
the $\mu^+$ precession frequency in the sample and $\omega_L$ is that in CaCO$_3$. The
data were corrected for a shape dependent macroscopic demagnetization
field according to the procedure described in Ref. [6]. The magnitude
of the Knight shift is small, on the order of 30 ppm, and negative.
The negative sign of the shift can be explained by the spin
polarization at the muon site. Although it was not possible to
determine an exact muon location in the lattice, the small value of
the Knight shift is consistent with the majority of previous
measurements on metal oxide compounds, where the $\mu^+$  is typically found
bound to a single oxygen ion by creating a "muoxyl" bond (O$\mu^+$)$^-$.\cite{7}

It has been established \cite{1,4} that the majority of carriers in
Li$_{0.9}$Mo$_6$O$_{17}$ are localized at molybdenum sites. Therefore, because of
the effective screening of Mo ions by oxygen ions, one can anticipate
a relatively low electronic spin density from those carriers on the
$\mu^+$.  This in turn leads to a value of $\omega_{\mu}$ that is close to $\omega_L$. Also, it
implies that, unlike in previously reported $\mu$SR results on $s=1/2$ AFM
chain compounds\cite{8}, the muon charge does not act as a strong
perturbation for the conducting chain. At the same time no significant
change has been observed in the muon Knight shift around 25 K.

In order to further investigate a possible magnetic origin of the
transition at 25 K, we performed a $\mu$SR experiment in a zero applied
magnetic field (ZF).  Figure 2 shows the results of this experiment.
As seen, there are only slow $1/T_1$ spin dynamics between RT and 4 K.
The measured relaxation rate is rather small ($\sim 0.2$ $\mu$s$^{-1}$), which is
characteristic of a fast fluctuating paramagnetic environment. The
small downturn above 200 K can be attributed to motional narrowing due
to rapid $\mu^+$ diffusion at elevated temperatures.

At the same time, the small change ($\sim10\%$) in the $\mu$SR
relaxation rate in the vicinity of $T_x=25$ K indicates that if a change in
the ground state occurs around 25 K, it most probably is attributable
to $\textit{increasing disorder}$ in the electronic environment to which the muon is coupled only
indirectly via the change in a local environment.
 \begin{figure}
     \centering
     \includegraphics[width=\columnwidth]{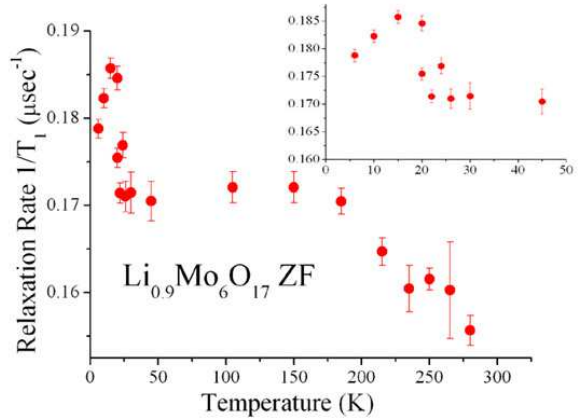}
     \caption{Measured $T-$dependence of the muon spin relaxation rate. 
 The inset shows the low temperature part around 25 K. } 
 \end{figure}
 
 This interpretation is  consistent
 with the most recently reported optical studies where 
 the role of disorder in the metal-insulator transition around 25 K
 it has been stressed .\cite{1}
 Moreover, absence of any periodic modulation in the measured ZF
 spectra means that $\textit{no SDW is detected}$ in the sample. Thus the CDW ground state 
 can be excluded. On the other
 hand, though less likely, one cannot exclude the possibility of very
 fast ($\geq 10^9$ s$^{-1}$) spin dynamics, which would be too fast to be observed
 on a $\mu$SR timescale.

We would like to express our gratitude for invaluable discussions to
I. Affleck and J. W. Allen and for the help with SQUID measurements to
E. Bruecher.



\end{document}